\documentstyle[aps,bezier,epsfig,amsmath,amssymb]{revtex}
         
\newcommand{\upd}{{\mathrm d}}
\renewcommand{\epsilon}{\varepsilon}

\begin{document}

\draft

\title{ \bf Is dynamic ultrametricity observable
in spin glasses?}

\author{ Ludovic BERTHIER$^{\star,\star\star}$,
Jean-Louis BARRAT$^\star$
and Jorge KURCHAN$^{\star\star\star}$}  

\address{$^\star$D\'epartement de Physique des Mat\'eriaux, 
Universit\'e C.
Bernard and CNRS, F-69622 Villeurbanne, France} 

\address{$^{\star\star}$Laboratoire de Physique, ENS-Lyon and CNRS,
 F-69007 Lyon, France }

\address{$^{\star\star\star}$PMMH, \'Ecole Sup\'erieure
de Physique et Chimie Industrielles,
F-75005 Paris, France}  

\date{\today}

\maketitle

\begin{abstract}
We investigate the dynamics of spin glasses from the `rheological'
point of view, in which aging is suppressed by the action of small,
non-conservative forces.
The different features can be expressed in terms of the scaling
of relaxation times with the magnitude of the driving force, which
plays the role of the  critical parameter. 
Stated in these terms, ultrametricity loses much of its mystery and
can  be checked rather easily.  This approach  also seems a natural
starting point to investigate what would be the  
real-space structures underlying 
 the hierarchy of time scales.
We study in detail the appearance of this many-scale behavior in a 
mean-field model, in which dynamic ultrametricity is clearly present.  
A similar analysis is
performed on  numerical results  obtained for a 3D spin glass:
In that case, our  results are compatible
with either that dynamic ultrametricity
is absent or that it develops so slowly
that even in experimental time-windows it is
still hardly observable.

\end{abstract}  

\pacs{LPENSL-TH-04/2000}   

% 05.70.Ln Non-equilibrium thermodynamics, irreversible processes 
% 75.10.Nr Spin-glass and other random models
% 75.40.Mg Numerical simulation studies      

\section{Introduction}

A remarkable feature of mean-field models of spin glasses
is the fact that pure states are organized in a hierarchical way.
Given three equilibrium configurations  (1,2,3) 
 they determine a  triangle whose sides are measured by the overlaps 
$q_{12}$,  $q_{13}$ and $q_{23}$.   If the configurations are chosen
 with the Gibbs weight then the two smallest overlaps are equal:
all triangles are isosceles.  
This property, valid in the thermodynamic limit,
defines an  ultrametric space \cite{book,mezard,ratovi}.

A first question to ask is whether  ultrametricity exists in
finite dimensional systems. Numerically, this is investigated by searching 
low temperature ground states of very small systems, and looking
for an ultrametric organization between them \cite{hart}.
Experimentally, however, spin glasses are always out of equilibrium,
and another relevant question may be to know what signatures 
of ultrametricity can be observable in a non-equilibrium system in the 
thermodynamic limit.
In a relaxing (aging) mean-field system, it has been shown
that ultrametricity would manifest itself in the 
dynamical behavior \cite{cuku,frme,cudo,review_aging}: 
the correlation functions  at three times 
$t_1 > t_2 > t_3 \gg 1$, $C(t_1,t_2)$,  $C(t_2,t_3)$ and $C(t_1,t_3)$ 
 satisfy the relation   
\begin{equation}
C(t_1,t_3)=\min \{C(t_1,t_2) \; , \; C(t_2,t_3) \}.
\label{eq1}
\end{equation}
This property is often termed `dynamic ultrametricity'.
A link between
dynamic and static ultrametricity has been established by Franz {\it et al}
\cite{franz},
who showed  that the existence of an ultrametric solution
from the dynamic point of view implies, under 
certain assumptions, the hierarchical
organization of pure states in  short-range systems.
However,  we are not aware of any report of dynamical ultrametricity
(in the form (\ref{eq1})) in
a simulated finite dimensional system, or in 
experiments \cite{review_aging,review_simu}.
The reason may be that
dynamic ultrametricity such as described by  (\ref{eq1})
is extremely difficult to observe numerically in an aging system
\cite{cit1}. 
Hence, for the 3D Edwards-Anderson model,
Franz and Ricci-Tersenghi \cite{frri} resorted 
to an indirect argument involving the coupling of replicas.
Their simulation (together with their argument) then  suggests a
situation for three dimensions similar to the one observed 
in mean-field models.
  
From the experimental point of view,  one obviously cannot
use the method of \cite{frri}. Moreover, noise autocorrelations are 
extremely difficult to measure, but one can check the version of
 Eq. (\ref{eq1}) involving the response functions \cite{cuku}:
\begin{equation}
M^{\mbox{\small{TRM}}}(t_1)\Big|_{t_w=t_3}= {\mbox{min}} \left\{
M^{\mbox{\small{TRM}}}(t_1)\Big|_{t_w=t_2}\; , \; 
M^{\mbox{\small{TRM}}}(t_2)\Big|_{t_w=t_3}
\right\},
\end{equation}
where $ M^{\mbox{\small{TRM}}}(t_a)|_{t_w=t_b}$ is the thermoremanent
magnetization at time $t_a$ when the field was cut off at time $t_b$.
{\em Such a relation is not observed experimentally}. Indeed, 
the fit
\begin{equation}
M^{\mbox{\small{TRM}}}(t_a)\Big|_{t_w=t_b} \simeq {\cal M}
\left(\frac{h(t_a)}{h(t_b)}\right)
\end{equation}
works very well for suitable functions $h(t)$ \cite{review_aging}, and
it is {\it incompatible}  with  dynamic ultrametricity.

As has been often remarked, glassy dynamics can be studied in two
complementary forms. One can study the relaxational aging system ---
and then
the large parameter is the waiting time $t_w$ --- or one can drive the
system in such a way it becomes {\em stationary} \cite{horner,cukupe,BBK}.
In this second `rheological' form, the relevant parameter is the intensity
of the driving force.  
Superconducting vortex systems with disorder have been studied
this way \cite{FFH,THPI}: in the presence of current the vortices are
driven by  a Lorentz force.
In a spin glass,
though less obvious to implement in experiments, 
the role of a  `stirring' force can be played by
non-symmetric couplings \cite{Grheso,ioma,mast}. 

Consider a system to which some non-conservative force of
strength $\epsilon$ is added. 
It is by now well established that in a quench experiment, 
after some transient, the system becomes 
stationary. 
If the system is glassy, the relaxation times in the
stationary regime
will diverge as $\epsilon \rightarrow 0$. (The time necessary to achieve   
stationarity is of the same order, and will diverge as well --- we
will not
study this in the paper.)
Indeed, we have that correlations decay for small $\epsilon$ as
$C(t)= C_{f}(t) + C_s(t,\epsilon)$,
where the subindices denote `fast' and `slow'.
[Recall that since we are in a stationary state, all the functions depend on 
a single time argument.]
$ C_{f}(t)$ represents
the `cage' motion, or the fluctuations inside a domain, and is not
affected
by a small perturbation. $C_s(t,\epsilon)$ is the slow motion, in
which
the system moves along the `almost flat' directions that are
responsible for aging ({\it e.g.} structural rearrangements in a glass,
domain wall motions in coarsening). 
The Edwards-Anderson parameter is given by
  $q \equiv \lim_{t \rightarrow \infty} 
\lim_{\epsilon \rightarrow 0} C(t)$.
A similar  decomposition holds for
$M^{\mbox{\small{TRM}}}(t)$.

The dynamic ultrametricity discussed above 
becomes extremely simple in terms of the
scaling
of $C_s(t,\epsilon)$ as  $\epsilon \rightarrow 0$.
One possibility is indeed  {\em simple scaling}:
\begin{equation}
C_s(t,\epsilon)= f \left(\frac{t}{{\bar t}(\epsilon)}\right)
\label{simple}
\end{equation}
where
${\bar t}(\epsilon) \rightarrow \infty$ when $\epsilon \rightarrow 0$,
and $f$ is a scaling function.
 Whenever Eq. (\ref{simple}) holds, we have that for two values of the
 correlation $C_1<C_2<q$:
\begin{equation}
\frac{t(C_1)}{t(C_2)} \underset{\epsilon \rightarrow 0}
{\longrightarrow} {\mbox{constant}},
\label{quot}
\end{equation}
where the relation $C_s(t)$ has been inverted to give $t(C)$.
The behavior of Eqs. (\ref{simple},\ref{quot}) is not the only possibility.
An alternative would be  that $C_s(t)$ is the sum of several
terms of the type (\ref{simple}), each having a different scaling
with $\epsilon$.
An extreme example of this is the {\em ultrametric law}, which for all
$C_1<C_2<q$ results in 
\begin{equation}
\frac{t(C_1)}{t(C_2)}  \underset{\epsilon \rightarrow 0}
{\longrightarrow}  \infty .
\label{ultrr}
\end{equation}
The relationship of this property to Eq. (\ref{eq1}) 
is immediate since, 
in the limit $\epsilon \rightarrow 0$, one has
$
C(t(C_1)+t(C_2)) \sim  C(t(C_1))= \min \{ C(t(C_1)), C(t(C_2) \},
$
which is the application of Eq. (\ref{eq1}) to a stationary situation.
This point is discussed in more detail in section \ref{calcul}.
The two following examples illustrate our discussion.
Simple scaling is, for small $\epsilon$: 
\begin{equation}
C_s(t) \sim f \left(
\frac{t}{ {\bar{t}}(\epsilon)}  \right) \quad \Leftrightarrow \quad 
t(C) \sim {\bar{t}}(\epsilon) j(C) ,
\end{equation}
where $j(C)$ is the inverse function of $f$, 
while dynamic ultrametricity may be obtained with the small $\epsilon$ scaling
\begin{equation}
\ln [t(C)] \sim {\bar{t}}(\epsilon) j(C)  \quad  \Leftrightarrow \quad 
C_s(t) \sim f \left(
\frac{\ln(t)}{ {\bar{t}}(\epsilon)}  \right),
\label{tauultr}
\end{equation} 
where  $ {\bar{t}}(\epsilon)$ may  diverge with  $\epsilon
\rightarrow 0$ as, for example, $ {\bar{t}}(\epsilon) \sim
\epsilon^{-a}$
and  $j(C)$ is a positive  decreasing function of $C$.   
Equation (\ref{tauultr}) clearly implies (\ref{ultrr}) and (\ref{eq1}), and 
expresses
the hierarchical scaling in a very
straightforward way.
In the stationary case, dynamic ultrametricity simply differs from simple
scaling in that to   make the 
correlation curves for different small $\epsilon$ collapse, {\em one has to
rescale $\ln(t)$ instead of $t$ with a function of $\epsilon$}.

Let us  emphasize that such a   simple
statement of ultrametricity is not possible in the aging case. 
Consider, as an example,
 an aging system with correlations evolving as 
$C(\tau+t_w,t_w)= \ln(t_w) / \ln(t_w+\tau)$, which is {\em
non-ultrametric} since $C(t_3,t_1)=C(t_3,t_2)*C(t_2,t_1)$
(as opposed to  Eq. (\ref{eq1})).
If we compare the time-differences $\tau(C)$
to reach two correlation values $C_1<C_2$,
it is easy to find:
\begin{equation}
\frac{\tau(C_1)}{\tau(C_2)} \sim (t_w)^{\frac{1}{C_1} -\frac{1}{C_2} }
\underset{t_w \rightarrow \infty}
{\longrightarrow} \infty.
\end{equation}
Hence, a criterion like (\ref{ultrr}) with $t_w$ playing the role of
the large parameter is {\em not} applicable
in the aging case, and one has to go back to Eq. (\ref{eq1}).

Our aims in this paper are the following:

(i) Within mean-field, the solution
exhibiting dynamical ultrametricity relies on an asymptotic analysis 
of the dynamical equations involving two-point correlation and response
functions, which may thus be questionable.
We will show that this analysis 
is indeed correct by solving numerically
the dynamical equations 
governing the driven dynamics of a mean-field spin 
glass in the stationary regime.
This is done on a very wide range of time scales, making us
confident that the asymptotic solution 
is the correct one.

(ii) Since ultrametricity should hold strictly in the asymptotic
limit of zero driving force (or equivalently in the infinite
waiting time limit in the aging case), it is important to study also
the preasymptotic regime to understand how ultrametricity gradually
develops, and how a full hierarchy of time scales appears.
We analyze then this preasymptotic regime in detail, 
thus characterizing the onset  of ultrametricity.

(iii) Having the mean-field dynamical behavior in mind, it
is very tempting to see if something similar
happens in a finite dimensional system. 
We have then performed  a Monte-Carlo simulation
of the 3D Edwards-Anderson model with asymmetrical couplings.
As is (unfortunately) usual in 3D spin glass simulations, our numerical
results may be interpreted in two different ways.
They are indeed compatible with an extremely slow appearance
of dynamic ultrametricity, {\it i.e.} much slower than in the mean-field case.
But an alternative view is that there is asymptotically only one relevant
time scale, or, in other words, 
that dynamic ultrametricity is not present at all.
In both cases the conclusion is that even if one assumes 
that ultrametricity is present, 
it has not fully developed for experimentally accessible time windows. 

The paper is organized as follows. In the next section,
the mean-field model under study is presented, and its preasymptotic
behavior is detailed in Section \ref{ultra}.
Section \ref{ea} presents our numerical results in three dimensions and
Section V contains our conclusions.

\section{A mean-field driven spin glass}
\label{model}

\subsection{Model}

We focus in the paper in a mean-field spin glass model
first introduced in Ref. \cite{theo}, where the statics was solved.
The  equilibrium dynamics was recently worked out \cite{cicr}.
It consists in a slight modification of the spherical $p$-spin model.
We consider indeed $N$ continuous variables $s_i$ ($i=1,\cdots,N$)
interacting through the Hamiltonian
\begin{equation}
H = -  \sum_{p=2}^{\infty} 
\sum_{j_1 < \cdots < j_p} J_{j_1 \cdots j_p} 
s_{j_1} \cdots s_{j_p}.
\end{equation}
In this expression, the $J$'s are random Gaussian variables,  symmetrical 
about the permutation of ($j_1,\cdots,j_p$) with mean zero and variance
\begin{equation}
 \overline{J_{j_1 \cdots j_p}^2} 
= \frac{p! {J_p}^2}{2 N^{p-1} }, 
\end{equation} 
so that the thermodynamic limit is well-defined.
A spherical constraint
$\sum_{i=1}^{N} s_i^2(t) = N$
is moreover imposed to the spins.
It is convenient to define also the function
\begin{equation}
g(C) \equiv \frac{1}{2} \sum_{p=2}^{\infty} {J_p}^2 C^p .
\end{equation}
The interesting case is when quadratic couplings are present together
with quartic and/or higher order interactions.
It was shown that the model belongs then to the universality
class of the Sherrington-Kirkpatrick model \cite{theo,cicr}:
it has a continuous transition between a paramagnetic phase
and a spin glass phase characterized by a full replica symmetry pattern
and a non-trivial probability distribution of overlaps $P(q)$.
Quantitatively, below the transition, one has
\begin{equation}
x(C<q) \equiv \int_0^C \upd q'P(q')=T \frac{g'''(C)}{2(g''(C))^{3/2}}; \quad
x(C>q)=1,
\label{parisi}
\end{equation}
where $q$ is the Edwards-Anderson parameter. 
This holds only if $J_2$ is big enough so that $x'(C)= P(C) > 0$.
We consider in the following the combination of second and sixth
order terms, keeping then only $J_2$ and $J_6$ different from zero.
In this particular case, one has $T_c=J_2$ (independent of $J_6$).
The positivity of  $x'(C)$ gives moreover
the inequality ${J_6}^2 < {J_2}^2 / 15$. 

\subsection {Driven dynamics}

To study the driven dynamics of the model, it is usual to consider
a Langevin equation
\begin{equation}
\frac{ \partial s_i (t)}{\partial t}
 = -\mu(t) s_i(t) - \frac {\delta H}
{\delta s_i(t)} +  \epsilon f_i^{\text{nc}}(t) 
+ \eta_i(t).
\end{equation}
As discussed in the introduction, we take for the (non-conservative)
driving force
\begin{equation}
 f_i^{\text{nc}} =  \sum_{\substack{j_1<\cdots<j_{k-1} \\
j_1,\cdots,j_{k-1} \neq i}} 
\tilde{J}_i^{j_1 \cdots j_{k-1}} s_{j_1} \cdots s_{j_{k-1}}.
\end{equation}
The parameter $\epsilon$ 
controls the strength of the force.
The couplings $\tilde{J}$'s in the 
driving force are random Gaussian variables, symmetrical about
the permutations of ($j_1,\cdots,j_{k-1}$), with mean zero and
variance
\begin{equation}
\overline{ \tilde{J}_i^{j_1 \cdots j_{k-1}} \tilde{J}_i^{ j_{1}
 \cdots j_{k-1} } } =   \frac{k!}{2 N^{k-1}}; \quad
\overline{ \tilde{J}_{i}^{j_1 \cdots j_{k-1}} \tilde{J}_{j_r}^{ j_1
 \cdots  i \cdots  j_{k-1} } } =  0.
\end{equation}
These couplings are partially
uncorrelated and contain thus an antisymmetrical part. 
This makes it impossible to write the driving force as the derivative
of an energy.
The parameter $\mu(t)$ ensures the spherical constraint and $\eta_i(t)$
($i=1,\cdots,N$) are random Gaussian variables with mean 0 and 
variance $2T$, where $T$ is the temperature of the heat bath.

The dynamics of the model is better
analyzed in terms of the autocorrelation
function $C(t,t') \equiv \sum_i \langle s_i(t) s_i(t') \rangle /N$ and
the response function $R(t,t') \equiv  \sum_i \langle \delta s_i(t) /
\delta \eta_i(t') \rangle /N$, since in the thermodynamic limit, 
$N \rightarrow \infty$, $C(t,t')$ and $R(t,t')$ verify
closed Dyson equations \cite{review_aging}.
The presence of the non-conservative force allows
to replace two-time functions $C(t,t')$ by single arguments 
functions $C(t-t')$, and the following equations
are obtained:
\begin{equation}
\begin{aligned}
\frac{\upd C(t)}{\upd t}  = & -\mu  C(t) + \int_{0}^{t} \upd t' 
\Sigma (t -t')  C(t')  + \int_{0}^{\infty} \upd t' 
\left[ \Sigma (t+t') C(t')  + D(t + t') 
R (t') \right], \\
\frac{\upd R(t)}{\upd t}  = & -\mu  R (t) + \int_{0}^{t} \upd t' 
\Sigma (t-t') R(t') ,\\
\mu  = &  T +  \int_{0}^{\infty} \upd t' \left[  D(t') 
R (t') + \Sigma (t') C(t') \right] ,\\
D(t)  \equiv & g'(C(t)) + 
\epsilon^2 \frac{k}{2} C(t)^{k-1}, \quad
\Sigma (t) \equiv  g''(C(t)) R(t).
\label{system}
\end{aligned}
\end{equation}
These integro-differential equations are associated with initial
conditions $C(0)=1$, $R(0^+)=1$ and with the condition $R(t<0)=0$ (causality).
The use of $C(-t)=C(t)$ in the derivation of Eqs. (\ref{system}) 
made these equations non-causal in the time difference, as can be seen
from the last integral in the equation for $C(t)$. 

\subsection{Asymptotic solution of the dynamical equations}
\label{calcul}

In order to make the paper self-contained, and
to set our notations, the asymptotic
analysis of Eqs. (\ref{system}) is now briefly recalled.
A more detailed computation can be found in \cite{cuku}.
`Asymptotic analysis' means that 
the limit $\epsilon \rightarrow 0$ is taken.

The first step of the analysis consists in making the decomposition
$C(t)=C_s(t)+C_f(t)$ and $R(t)=R_s(t)+R_f(t)$ between
a fast and a slow part,
and to derive
the equations verified by each part.
The equations for $C_f$ and $R_f$ are solved by making the ansatz
that they satisfy the fluctuation-dissipation theorem (FDT)
$ T R_f(t)=-\upd C_f(t) / \upd t$.
Taking the limit $t\rightarrow \infty$ gives then the following relation:
\begin{equation}
\mu + \frac{g'(q)-g'(1)}{T} = \frac{T}{1-q},
\label{fausse}
\end{equation}
where $q \equiv \lim_{t \rightarrow \infty} \lim_{\epsilon \rightarrow 0}
C(t)$ is the Edwards-Anderson parameter.
The slow parts verify:
\begin{equation}
\begin{aligned}
\frac{\upd C_s(t)}{\upd t}  = & -\mu  C_s(t) + \int_{0}^{t} \upd t' 
\Sigma_s (t -t')  C_s(t')  + \int_{0}^{\infty} \upd t' 
\left[ \Sigma_s (t+t') C_s(t') + D_s(t + t') 
R_s (t') \right] \\
& + \frac{g'(1)-g'(q)}{T} C_s(t) + \frac{1-q}{T} D_s(t)  
, \\
\frac{\upd R_s(t)}{\upd t}  = & -\mu  R_s (t) + \int_{0}^{t} \upd t' 
\Sigma_s (t-t') R_s(t') + \frac{g'(1)-g'(q)}{T} R_s(t)
 +  \frac{1-q}{T} \Sigma_s(t) 
 ,\\
\mu  = &  T +  \int_{0}^{\infty} \upd t' \left[  D_s(t') 
R_s (t') + \Sigma_s (t') C_s(t') \right] + \frac{g'(1)-qg'(q)}{T}
 ,\\
D_s(t)  \equiv & g'(C_s(t)), \quad
\Sigma_s (t) \equiv  g''(C_s(t)) R_s(t).
\label{system_lent}
\end{aligned}
\end{equation}

The second step stems from the observation that once the first order
derivatives are dropped out (which is of course justified
in the slow regime), the above equations become 
invariant under a reparametrization of time $t \rightarrow h(t)$.
This suggests that $C_s$ and $R_s$ are related 
through a reparametrization-invariant formula, namely
\begin{equation}
 R_s(t) = - \frac{X(C_s(t))}{T}\frac{ \upd C_s(t)}{ \upd t}.
\end{equation}
This amounts to an extension of   the FDT to this non-equilibrium
situation \cite{cukupe} by the introduction of an effective temperature
$T_{\text{eff}}(C) \equiv T/X(C)$.
In the same spirit, the function $f$ defining `triangles'
is introduced: 
\begin{equation}
C_s(t_1+t_2) \equiv f(C_s(t_1),C_s(t_2)),
\label{f}
\end{equation}
Let us emphasize that the main assumption of the analysis is 
that the functions
$X(C_s,\epsilon)$ and $f(C_1,C_2,\epsilon)$ have a continuous limit
when $\epsilon \rightarrow 0$.
It is also convenient to define 
the function $\bar{f}$: 
$C_s(t_1) \equiv \bar{f}(C_s(t_2),C_s(t_1+t_2))$.
This allows to rewrite Eqs. (\ref{system_lent})
in such a way that the time disappears:
\begin{equation}
\begin{aligned}
0 = & -\frac{T}{1-q} C_s + \frac{1-q}{T} g'(C_s) + 
\frac{1}{T} \int_{C_s}^q \upd C_s' 
g''(C_s')X(C_s')\bar{f} (C_s',C_s) \\
& + \frac{1}{T} \int_0^{C_s} \upd C_s' X(C_s')g''(C_s')
\bar{f}(C_s,C_s') -\frac{1}{T}
\int_0^{C_s} \upd C_s' g''(C_s') F[\bar{f}(C_s,C_s')] ,\\
0= & - \frac{T}{1-q} F[C_s] + \frac{1-q}{T} H[C_s] + \frac{1}{T} 
\int_{C_s}^q \upd C_s' g''(C_s') X(C_s') F[\bar{f}(C_s',C_s)] ,\\
F[C_s]  \equiv & - \int_{C_s}^q \upd C_s' X(C_s'),\quad
H[C_s] \equiv  - \int_{C_s}^q \upd C_s' X(C_s') g''(C_s'),
\label{notime}
\end{aligned}
\end{equation}
where
Eq. (\ref{fausse}) has been used.

The last step consists in computing explicitly $X$ and $f$
from Eqs. (\ref{notime}).
This is done by introducing 
`fixed points' of $f$ \cite{cuku}.
A fixed point  $q^\star$ of $f$ satisfies
\begin{equation}
f(q^\star,q^\star)= q^\star.
\end{equation}
In the stationary  context we are discussing, the physical meaning 
of such values of the correlation is very clear, 
since from the definition
of $f$,
\begin{equation}
q^\star = C_s(t) \Rightarrow
C_s(2t)=f(C_s(t),C_s(t))=C_s(t),
\end{equation}
which simply means that $q^\star$ is a {\it plateau} in the 
correlation function.
Two trivial such fixed points are $q=0$ and $q=1$.
It is then rather straightforward \cite{cuku} to solve Eqs. (\ref{notime}).
This gives $X(C)$
for $C \in [0,q]$, 
\begin{equation}
X(C) = T \frac{g'''(C)}{2(g''(C))^{3/2}},
\label{FDR}
\end{equation} 
and
the matching with the FDT regime ($X(C>q)=1$) determines $q$ which 
satisfies $T=(1-q)\sqrt{g''(q)}$.
The fact that $X(C)$ coincides with $x(C)$ (Eq. (\ref{parisi}))
is a remarkable property of this class of mean-field 
spin glass models \cite{cuku,franz}.
The properties of $f$ are also obtained.
It is first shown that $\forall \, C \in [0,q]$, $f(C,C)=C$,
and there is hence a continuum of fixed points.
Crudely speaking, it can be said that, in the asymptotic limit,
each value of $C<q$ corresponds to a plateau value in the correlation function
(in the sense that the correlation will not decay below 
this value in a finite time).
This  ultimately means
that we will have a behavior as described by Eq. (\ref{ultrr}).
It is also shown that
\begin{equation}
\forall \, (C_1,C_2) \in [0,1]^2 \smallsetminus 
 [q,1]^2 \quad f(C_1,C_2) = \min \{ C_1,C_2 \},
\label{ultraeq}
\end{equation}
which is the ultrametric relation, as presented in the introduction.

The conclusion of this section, is that, apart from the trivial
short time behavior where $C(t) > q$ and FDT holds $X(C>q)=1$,
the relaxation below $q$ is characterized by a hierarchy of
time scales and a non-trivial FDT $X(C<q) < 1$.

\section{Ultrametricity from dynamical correlations}
\label{ultra}

As stated in the preceding section, the asymptotic
solution of the dynamical equations exhibits a `many-plateau pattern', which
is an unusual feature.
To see how this asymptotic solution is approached,
Eqs. (\ref{system}) were solved numerically, in
the particular case where $g(C)=  C^2/2 + C^6/30$,
and $k=2$.
We shall mainly work at $T=0.25=0.25T_c$, 
where the Edwards-Anderson parameter
is $q \simeq 0.787$.
To solve these equations, a combination of numerical methods of Refs.
\cite{ng,arnulf} has been used.

\subsection{Hierarchy of time scales}
\label{hierarchy}

\begin{figure}
\begin{center}
\psfig{file=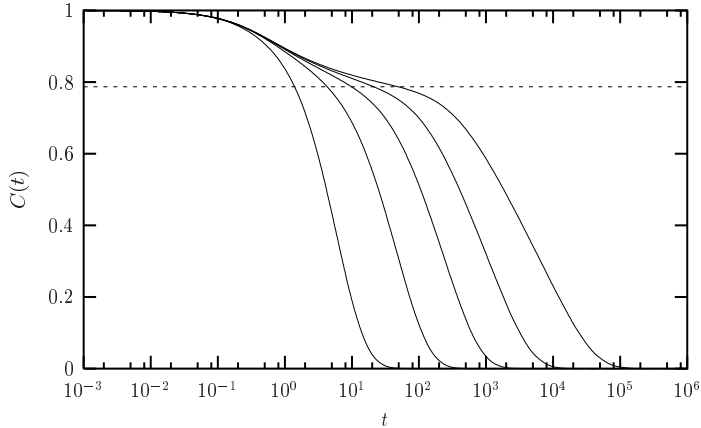,width=10cm,height=6.5cm} 
\caption{Correlation function as a function of time for different values
of the asymmetry for the mean-field model, at $T=0.25$.
From left to right, $\epsilon=2.25$, 0.8, 0.5, 0.35, 0.248.
The horizontal dashed line is the value of the
Edwards-Anderson parameter $q\simeq 0.787$.}
\label{correlation}
\end{center}
\end{figure}

A first element that is missed by the above analysis is  
the functional form 
of the correlation functions.
They are depicted in 
Fig. \ref{correlation}, for different values of $\epsilon$
and for $T=0.25$.
As expected,
two different regimes are clearly present.
For $C(t)>q$, the relaxation depends very weakly on the asymmetry, while
for $C(t)<q$, the smaller the asymmetry, the slower
the relaxation.

\begin{figure}
\begin{center}
\psfig{file=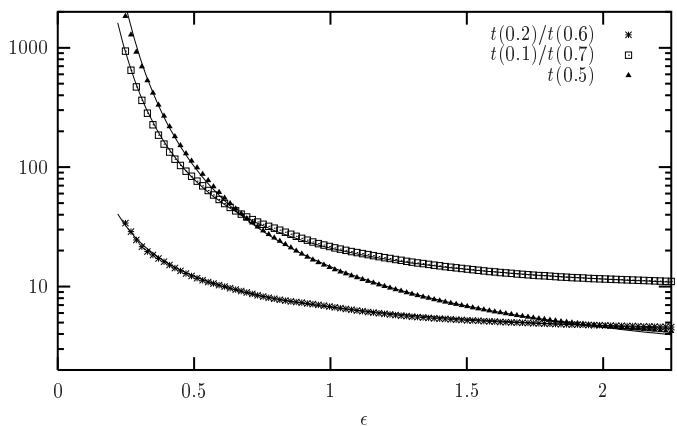,width=10cm,height=6.5cm} 
\caption{Two different ratios of relaxation times
as a function of the asymmetry for the mean-field model.
Note that the vertical axis is in a logarithmic scale.
Also plotted is the divergence of $t(0.5)$. 
The lines are fits of the form (\ref{numresult}), with $j(C)$ as a fitting 
parameter.}
\label{times}
\end{center}
\end{figure}

The main question that cannot be answered analytically
is the precise dependence of the relaxation times
on the parameter $\epsilon$ that controls the strength
of the asymmetry. The `many-plateau pattern' discussed above
means that, in the preasymptotic regime (non-zero $\epsilon$),
the correlation will stay a large, albeit finite time around the same value.
This characteristic time increases with decreasing $C$, as expressed by
Eq. (\ref{ultrr}).
In order to test this separation of time scales, we compute 
the ratio of relaxation times for two fixed values $C_1$ and $C_2$ of
 the correlation as a function of $\epsilon$.
Two such ratios are represented in Fig. \ref{times}, where it is clearly seen
that they indeed diverge in the small asymmetry limit. 
A consequence is  that the relaxation cannot be represented
by a single time scale, but involves a full hierarchy of them.
In particular, a stretched  exponential fit such as the one 
 proposed in \cite{ioma} for
the Sherrington-Kirkpatrick model with asymmetry can only be approximate.

The numerical solution of the dynamical equations suggests
the following dependence for $C<q$:
\begin{equation}
t(C) \sim \exp {\Big [}  {\bar t}(\epsilon)  j(C)  {\Big ]}; 
\quad {\bar t}(\epsilon) \sim  \epsilon^{-0.65}.
\label{numresult}
\end{equation}
In this expression, 
$j(C)$ is a positive decreasing function of the correlation.
It plays the same role as in the example of the 
introduction, Eq. (\ref{tauultr}).
Numerically, $j(C)$ is consistent with a linear variation:
$j(C) \simeq j(C=0) - b C$.
For a fixed value $C$ of the correlation, $t(C)$ grows faster than
a power law of $\epsilon$, as was noted in Ref.\cite{ioma}.
It also follows that
\begin{equation}
\frac{t(C_1)}{t(C_2)} \sim \exp {\Big [} {\Big (} j(C_1)-j(C_2)
{\Big )} {\bar t}
(\epsilon)  {\Big ]},
\end{equation}
which means that the ratios of two time scales may be fitted
by the same functional form (\ref{numresult}) as the time scales themselves.
This is also displayed in Fig. \ref{times}.

Let us note that the scaling (\ref{numresult}) is very reminiscent of the
`creep' regime scaling for vortex glasses \cite{FFH}, with the role of
$\epsilon$ played by the current and $C$ a measure of the average
squared transverse displacements along the vortex. 

\begin{figure}
\begin{center}
\psfig{file=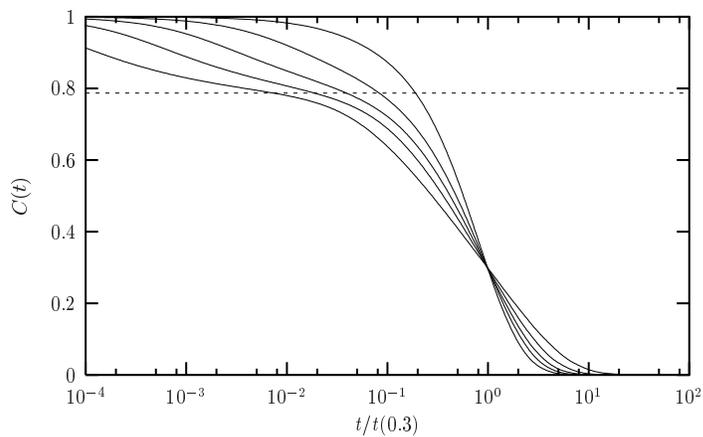,width=10cm,height=6.5cm} 
\caption{
Correlation function for the same values of the asymmetry as in 
Fig. \ref{correlation}. The time is rescaled by $t(0.3)$ so that curves 
superpose for $C=0.3$. Away from this value, there is no collapse for
small $\epsilon$.
}
\label{rescale}
\end{center}
\end{figure}

\begin{figure}
\begin{center}
\psfig{file=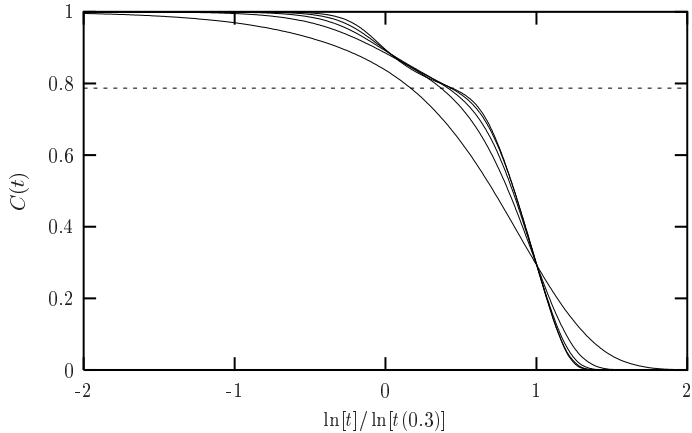,width=10cm,height=6.5cm}
\caption{
Correlation function for the same values of the asymmetry as in 
Fig. \ref{rescale}, but here the {\em logarithm} of time is rescaled
by $\ln[t(0.3)]$. The collapse for small $\epsilon$ is good.
}
\label{resc}
\end{center}
\end{figure}

The presence of this hierarchy of time scales in (\ref{numresult}) implies
that the correlation curves
can not be superimposed by rescaling the time.
This is illustrated in Fig. \ref{rescale}, where the time is rescaled 
so that the curves meet at the value $C=0.3$.
It can be seen that
the curves are more and more horizontal around $C=0.3$, when the asymmetry is
decreased. 
On the  contrary, if we rescale the logarithm of the time to make the
curves meet at  $C=0.3$ (Fig. \ref{resc}) we find that the curves
tend to collapse. In other words: {\em shifting} $\ln(t)$ (as in simple
scaling) does not make the curves collapse, while {\em stretching} 
$\ln(t)$ does.
A similar picture would have been obtained by choosing
any value $C\in[0,q]$.

This rescaling (or absence thereof)
of the correlation functions in the driven 
dynamics of a spin glass is a direct 
and very simple test of dynamic ultrametricity. 
Let us emphasize again that, as already mentioned in the introduction,
this is not true for the aging regime.

\subsection{Ultrametric relation and FDT}

\begin{figure}
\begin{center}
\psfig{file=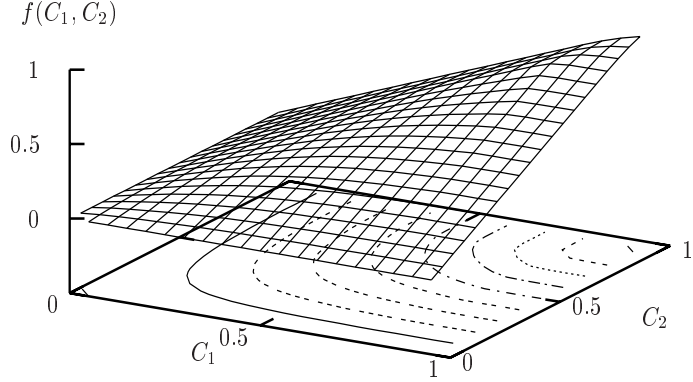,width=10cm,height=6.5cm} 
\caption{Function $f$ (Eq. (\ref{f})), for the value $\epsilon=0.248$ in 
the mean-field model.
The curves shown in the $(C_1,C_2)$ plane are projections
of horizontal cuts through the surface for $f=0.1,0.2,\cdots,0.9$.}
\label{surf}
\end{center}
\end{figure}

The function $f(C_1,C_2)$ introduced in Eq. (\ref{f}) 
satisfies in the asymptotic limit the
ultrametric relation (\ref{ultraeq}).
It is then natural to try to understand its preasymptotic
behavior. 
A three dimensional
view of this two-variable function
is given in Fig. \ref{surf}, for $\epsilon=0.248$ (the slowest
relaxation in Fig. \ref{correlation}).
Also plotted in the plane ($C_1,C_2$) are the constant-$f$ contours.
These contours would be right angles in the limit of
vanishing asymmetry.

\begin{figure}
\begin{center}
\psfig{file=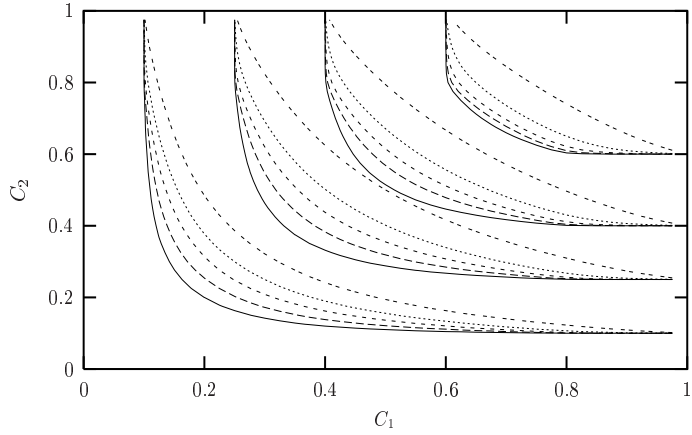,width=10cm,height=6.5cm}
\caption{
Evolution of the projections 
for $f=0.1$, 0.25, 0.4 and 0.6 (see text)
in the plane $(C_1,C_2)$
for the values of the asymmetry
as in Fig. \ref{correlation}: $\epsilon=2.25$, 0.8, 0.5, 0.35 and 0.248.
At each level $f$, the upper curve is for the larger asymmetry,
and decreasing the asymmetry make the curves move towards right angles.} 
\label{iso}
\end{center}
\end{figure}

To see how the function $f$ evolves towards its asymptotic value
(the ultrametric relation (\ref{ultraeq})),
the evolution of the contours for different $\epsilon$
is represented in Fig. \ref{iso}.
These curves are very clearly evolving towards right angles,
as expected from the analysis of the preceding section.

\begin{figure}
\begin{center}
\psfig{file=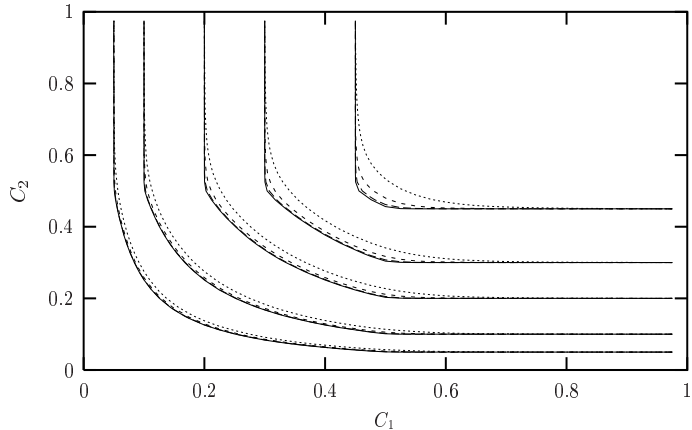,width=10cm,height=6.5cm}
\caption {Evolution of the projections 
for $f=0.05$, 0.1, 0.2, 0.3 and 0.45 (see text)
in the plane $(C_1,C_2)$
for the $p$-spin model (here $p=3$) studied in Ref. [15]. 
At each  level $f$, the upper curve is for the larger asymmetry.
In this case, the Edwards-Anderson parameter is $q \simeq 0.501$.
The asymmetry (4 values are represented)
is chosen so that correlation functions 
cover a range of time scales similar to that of Fig. \ref{correlation}.
When the asymmetry is decreased, 
the function evolves rapidly towards its limiting
(non-ultrametric) shape.}
\label{p=3}
\end{center}
\end{figure}

It is interesting to compare these curves with the ones
obtained from Ref. \cite{BBK} 
for a system ($p$-spin with $p=3$) with a single time scale, 
{\it i.e.} without ultrametricity.
The result is shown in Fig. \ref{p=3}, for correlations that 
evolve on a similar range of time scales to make the comparison
relevant.
In this case, the asymptotic
analysis reveals that there exists asymptotically a function $f$
defined as above, but
it is not given by the ultrametric relation (\ref{ultraeq}).
The difference between the two systems is very clear 
from the
comparison of Figs. \ref{iso} and \ref{p=3}: in the 
latter, the function $f$ rapidly 
saturates to its asymptotic (non-ultrametric) value. 

\begin{figure}
\begin{center}
\psfig{file=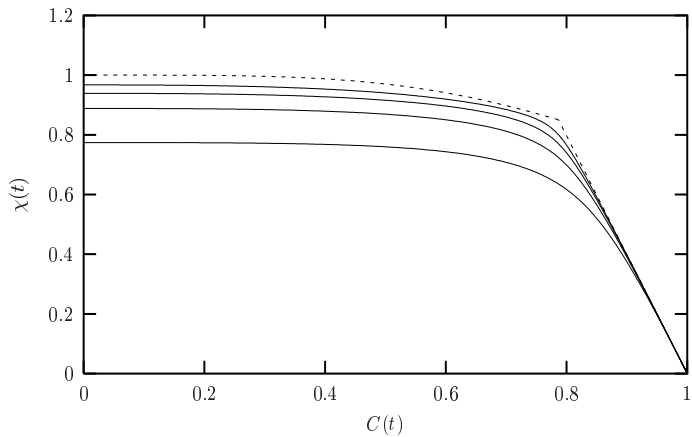,width=10cm,height=6.5cm}
\caption{
Parametric plot of the susceptibility {\it vs} the correlation functions
for $T=0.25$  and
$\epsilon=0.8$, 0.5, 0.35 and 0.248 for the mean-field model.
The dashed curve is the analytic result, Eq. (\ref{FDR}), for the limit 
$\epsilon \rightarrow 0$.}
\label{parametric}
\end{center}
\end{figure}

To make the analysis of the mean-field dynamics complete,
the usual plot \cite{cuku} of the integrated response function
$\chi(t) \equiv \int_0^t \upd t' R(t')$ as a function
of the correlation function $C(t)$, parameterized by the time $t$,
is done in Fig. \ref{parametric}.
As expected from the above analytical results, the FDT holds
for $C(t)>q$ and it is strongly violated for smaller values
of the correlation.
It is clear that in the limit of zero asymmetry, the
analytic expression for the function $X(C)$ that generalizes
the FDT to non-equilibrium situations
will be recovered.
The fact that the same limiting FDT violations happen
in gently driven 
and aging systems has been suggested in Ref. \cite{cukupe},
and we have verified here that this (mean-field) 
result holds also for systems with many time scales.

\section{Simulation of the 3D Edwards-Anderson model}
\label{ea}

We turn now
to the numerical results obtained for
a 3D spin glass model, the results
of the preceding sections being a guide 
to investigate its stationary driven dynamics.
We  use the same notations for quantities that play a similar role
in the simulation and in the mean-field model,
the distinction between the two cases being clear from the context.

\subsection{Model and details of the simulation}
The model under study is defined through its Hamiltonian 
\begin{equation}
H = - \sum_{\langle i,j \rangle} J_{ij} s_i s_j,
\end{equation}
where $s_i$ ($i=1,\dots,N$) are $N=L^3$ Ising spins 
located on the sites of 3D cubic lattice of linear size $L$,
with periodic boundary conditions.
The sum $\langle i,j \rangle$ runs over nearest neighbors
and the $J$'s are chosen randomly from a bimodal distribution
$J_{ij}= \pm 1$.
The model has been extensively studied \cite{review_simu}:
it exhibits a second order phase transition at the 
critical temperature $T_c = 1.11 \pm 0.04$ from a paramagnetic to a
spin glass phase \cite{kayo}.

To drive the system, a coupling $\tilde{J}_{ij}$ is added on each link.
The $\tilde{J}$'s are chosen from a bimodal
distribution $\tilde{J}_{ij} = \pm \epsilon$, and are antisymmetrical:
$\tilde{J}_{ij} = - \tilde{J}_{ji}$.
The small parameter $\epsilon$ controls the strength 
of the driving force.
The spins are randomly sequentially updated through a standard 
Metropolis algorithm, and one Monte Carlo step
represents $N$ attempts to update a spin.
Numerical results are presented for a linear size $L=20$ ($N=8000$ spins), 
where
finite size effects are negligible for the time scales investigated here 
\cite{ogielski}.

The temperature has been chosen so that the Edwards-Anderson
parameter is comparable to the one of the mean-field case studied
before.
At
$T=0.6 \simeq 0.54 T_c$, the Edwards-Anderson parameter
was roughly
estimated in an off-equilibrium simulation (with $\epsilon = 0$)
through its dynamical definition $q= \lim_{t\rightarrow \infty} \lim_{t_w 
\rightarrow \infty} C(t,t_w) \sim 0.8$ (we had $q \simeq 0.787$ in 
the mean-field study).
Similarly, 
5 different values of the asymmetry were studied:
$\epsilon = 0.5$, 0.4, 0.3, 0.25 and
0.2, so that the range of time scales is comparable to the previous
mean-field results.
Remarkably, initial conditions are 
irrelevant, since a stationary state is reached after
a time which depends on the intensity of the drive.
As was already noted in Ref. \cite{BB},
this is an unusual feature for glassy systems 
where cooling procedures are known to be crucial.
Stationarity allows moreover to average the correlation functions
over different initial times, so that very few 
averages over the disorder have been necessary (typically 5), provided
the first steps of the simulation are discarded. Stationarity has been
carefully checked throughout the simulation.

\subsection{Looking for ultrametricity}

The dynamical quantity of interest is the spin-spin autocorrelation 
function, which in the TTI regime reads
\begin{equation}
C(t) \equiv \frac{1}{N} \sum_{i=1}^N 
\langle \overline{s_i(t+t_0) s_i(t_0)} \rangle.
\end{equation}
The overline means that an average over  disorder is performed, 
while $\langle \cdots \rangle$  stands for an average 
over different initial times $t_0$, all chosen
in the TTI regime.

\begin{figure}
\begin{center}
\psfig{file=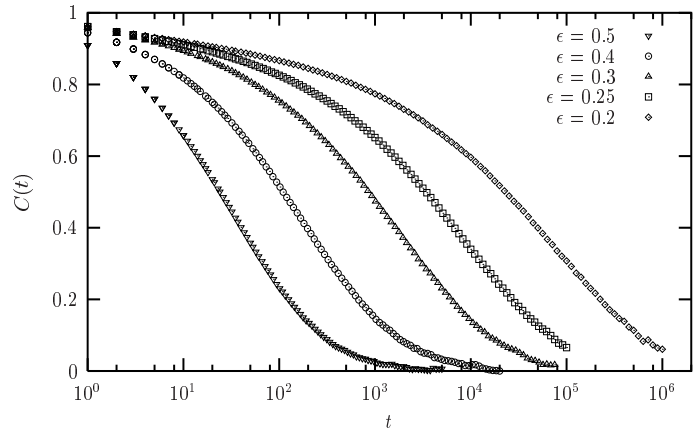,width=10cm,height=6.5cm}
\caption{
Correlation functions for the 3D Edwards-Anderson model for different
values of the asymmetry at $T=0.6$.}
\label{ea_corr}
\end{center}
\end{figure}

The autocorrelation functions for different values of the asymmetry
at $T=0.6$ are represented in Fig. \ref{ea_corr}.
As for the mean-field case, two different regimes are present.
The short time relaxation towards $q \sim 0.8$ is very weakly affected 
by the driving force, while the time to relax towards 0
dramatically increases when $\epsilon$ is lowered.

\begin{figure}
\begin{center}
\psfig{file=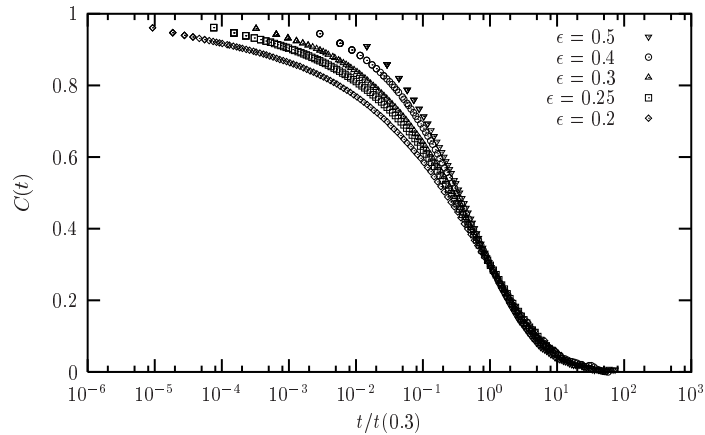,width=10cm,height=6.5cm}
\caption{Rescaling of the correlation functions of the Fig. \ref{ea_corr}.
The rescaled time is such that all the curves meet at the value $C=0.3$.
This figure has to be compared with Fig. \ref{rescale}.}
\label{ea_collapse}
\end{center}
\end{figure}

The crucial point is to look for the possible presence of
the ultrametric relaxation pattern described in the preceding sections.
We have seen above that a simple test
of the presence of ultrametricity is the rescaling of the time.
The same rescaling that was done in Fig. \ref{rescale} for the
mean-field case is now performed
for the 3D case in Fig. \ref{ea_collapse}.
Although not perfect, this rescaling
works remarkably well for the smallest values of $C$.
It is important to note that even if ultrametricity is absent,
the values of correlation  $1<C \lesssim q$ (which
do not depend on the asymmetry) will not scale together with
the $C<q$ portion of the curves. 
This may be a source of errors if the value
of $q$ is unknown. Thus,
preasymptotic effects affect the quality of the rescaling
in the region $C \sim q$, since it can be seen from Fig. \ref{ea_corr}
that the plateau is not completely developed.
All these points may explain why the rescaling is good only
for the smallest values of $C$.

However, 
a very slow flattening of the curves around the value $C=0.3$
cannot be completely excluded from this figure.
But it is very clear that if there is a separation of
time scales, then it {\it will become evident only if a
much larger time window is analyzed}.

\begin{figure}
\begin{center}
\psfig{file=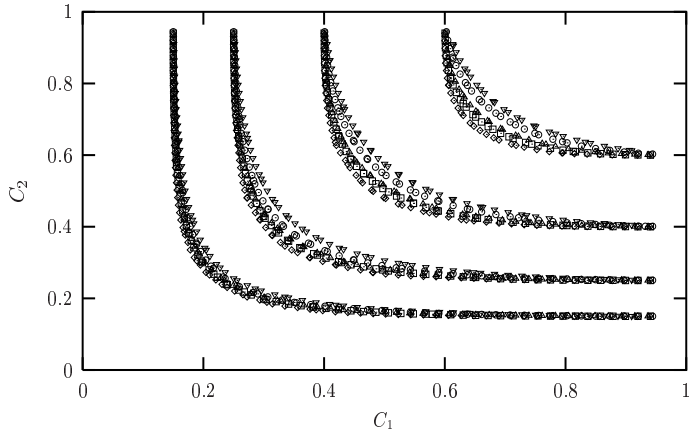,width=10cm,height=6.5cm}
\caption{Test of the function $f$ for the 3D Edwards-Anderson model
for $f=0.6$, 0.4, 0.25 and 0.15.
The values of the asymmetry are the same as 
in Figs. \ref{ea_corr} and \ref{ea_collapse}, 
and are represented with the same symbols.
This curve has to be compared with Figs. \ref{iso} and \ref{p=3}.}
\label{ea_leti}
\end{center}
\end{figure}

The second interesting quantity to study is the function $f(C_1,C_2)$
introduced in Eq. (\ref{f}).
Its evolution for different values of the 
asymmetry is depicted in Fig. \ref{ea_leti}, 
with the same construction as for Figs. \ref{iso} and \ref{p=3}.
The function $f$ indeed evolves, but
very slowly when compared
to the mean-field case. 
The quantitative behavior of $f$ for the 3D model is 
more reminiscent of the $p=3$ case 
than of the mean-field spin glass case.
Once again, this figure does not allow to decide clearly between
a limiting smooth -- as it is for the $p=3$ case -- 
or ultrametric $f$-function, because interesting things 
may happen on time scales that are inaccessible in our 
simulation time.

In our opinion, the important point that clearly emerges
from the figures is that,
on a given time window (we have 4 decades here), 
the relaxation does not appear 
to be typical of an ultrametric system and 
a single-time-scale description is very accurate.
This may explain why dynamic ultrametricity
has not been observed in aging experiments, which
span some six decades $\sim$  (100 Hz. - 10 hours).

\subsection{FDT}

\begin{figure}
\begin{center}
\psfig{file=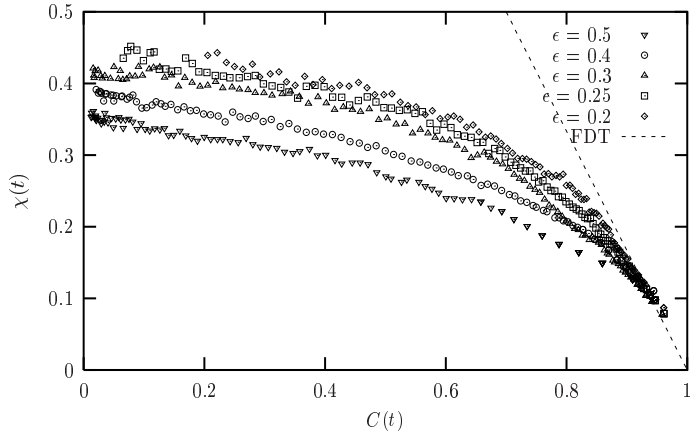,width=10cm,height=6.5cm}
\caption{Parametric plot of the susceptibility
vs the correlation functions for the 3D model for
 different values of the asymmetry, at $T=0.6$.
The curves saturate, in the small asymmetry limit, to a smooth limiting
curve.}
\label{ea_chi}
\end{center}
\end{figure}

We have also investigated the way the FDT is violated
in the driven 3D Edwards-Anderson
model, since the mean-field theory suggests
that these violations are the same as for an aging system.
To our knowledge, there are no numerical confirmation of 
this prediction for spin glasses in the literature.
For this purpose,
correlation functions have to be compared to susceptibility
curves.
The correlation functions are taken from Fig. \ref{ea_corr}.
The way of computing the susceptibility is now standard \cite{3D}.
A stationary magnetic field $h_i$ is applied in each site at time $t_0$.
It is random in space, and we take it from a Gaussian
distribution with mean 0 and variance $ \overline{h_i h_j} = h_0^2$.
The staggered magnetization 
\begin{equation}
m(t) \equiv \frac{1}{h_0 N} \sum_{i=1}^{N} h_i s_i(t),
\end{equation}
is recorded 
for all $t>t_0$. 
In the linear response regime (we work with $h_0=0.1$, as in previous studies
\cite{review_simu,3D}), the susceptibility
is obtained from $m(t)$ as
$\chi(t) \simeq m(t)/ h_0$.
The results are averaged over 
several (from 50 to 300) realizations
of the field, and over different initial times $t_0$.
They are presented in Fig. \ref{ea_chi}.

The results for the violations of the FDT are clearly
very similar to  the mean-field ones.
These curves exhibit two different regimes: for short times ($C \gtrsim q$),
the FDT is well satisfied, whereas for longer times ($C \lesssim q$),
it is violated.
The curves also clearly saturate to a smooth limiting curve in the 
small asymmetry limit.
The shape of this limiting curve 
is compatible with the limiting non-trivial
parametric curves that have already been found in simulations
of 3D spin glasses in the aging regime \cite{3D}.
The fact that both situations (aging and driven dynamics) exhibit
the same kind of FDT violations deepens, in our opinion,
the physical meaning of the quantity $X(C)$ 
(the so-called fluctuation-dissipation ratio) that can 
be extracted
from this plot \cite{franz,cukupe,frvi}.

\section{Conclusions}

We have studied in this paper the behavior of spin glasses
in a `rheological' setting, in which the dynamics is stationary and
the control parameter is the strength of the driving force.

We have checked the asymptotic analysis \cite{cuku,frme,cudo} 
for mean-field models
 through the numerical integration of the equations
governing the dynamics.
This has confirmed  the presence
of a full hierarchy of time scales in the relaxation of the
correlation,
 Eq. (\ref{ultraeq}), and a
`many-time-scale, many-temperature scenario' \cite{BBK}.
This scenario can be seen as the dynamic counterpart of static ultrametricity.
Simple scaling time dependencies, such as assumed in Ref.
\cite{ioma} could  be justified for mean field models
only  as an 
approximation valid in some time-window, but do not hold strictly in
the large time limit.
Previous numerical studies of the aging dynamics
of the Sherrington-Kirkpatrick model had 
pointed out the lack of simple $t/t_w$-scaling \cite{maparo},
but did not investigate the presence of ultrametricity.
Since there exist now powerful algorithms to solve dynamical 
two-time equations for large times \cite{arnulf_horner}, 
it would be interesting to have an analysis of the aging 
dynamics of a mean-field ultrametric model following the lines 
introduced in this paper.

For the 3D spin glass, our simulation suggests that
the  stationary  driven dynamics is accurately
 interpreted within a single time scale
relaxation pattern. We cannot
decide whether our results imply an extremely slow
appearance of ultrametricity (such that, even at experimental times,
 it is not
fully observable), or its absence.

Whether transient or permanent,
the single time scale relaxation obtained in  our 3D simulation is
in complete accordance with all the known numerical 
and experimental studies of the aging regime of 3D spin glasses 
\cite{review_aging,review_simu,heiko}.
In all known cases, two-time functions 
are indeed very well approximated by
$M(t,t_w) \simeq {\cal M} (t/t_w)$,
in contradiction with the dynamical mean-field theory, as 
we emphasized all along the paper.
It is interesting to note that the version of the 
`droplet theory' \cite{fisher}
which predicts the scaling 
$C(t,t_w) \sim {\cal C} (\ln(t)/\ln(t_w))$
fails also in reproducing experiments and simulations, since the
logarithmic law
is too slow.

Numerical results on 4D spin glasses also show 
that correlation functions are rather well represented by 
a $t/t_w$-scaling in the aging regime \cite{ricci}, 
as is found in 3D. 
It would be extremely interesting to
go back to the  4D simulations with asymmetrical couplings \cite{mast} to
study precisely the stationary  regime,
and see if, like in our simulation, there is a single time scale.
It would be a disappointing result if 
all the richness of the  ultrametric construction
in spin glasses were a pure
$D \rightarrow \infty$ feature. 

The situation is made even more puzzling
by the fact that this construction and the associated  separation
of time scales are often invoked to explain the results of 
temperature cycling experiments \cite{review_aging,cuku2}.
This is perhaps related to the difficulty in reproducing these
experimental results with the 3D EA model with 
numerical simulations \cite{felixcm}.


\begin{references}
\bibitem{cit1} As far as we know, this was only attempted in
  \cite{cuku}, with rather inconclusive results.
\end{references}
\end{document}